# Real Time Discovery of Dense Clusters in Highly Dynamic Graphs: Identifying Real World Events in Highly Dynamic Environments

Manoj K Agarwal[*]
IBM Research-India, New Delhi
manojkag@in.ibm.com

Krithi Ramamritham
IIT-Bombay, Mumbai, India
krithi@cse.iitb.ac.in

Manish Bhide
IBM India Software Labs, Hyderabad
abmanish@in.ibm.com

## ABSTRACT

Due to their *real time* nature, microblog streams are a rich source of dynamic information, for example, about *emerging events*. Existing techniques for discovering such events from a microblog stream in real time (such as Twitter trending topics), have several lacunae when used for discovering emerging events; extant graph based event detection techniques are not practical in microblog settings due to their complexity; and conventional techniques, which have been developed for blogs, web-pages, etc., involving the use of keyword search, are only useful for finding information about *known* events. Hence, in this paper, we present techniques to discover events that are unraveling in microblog message streams in real time so that such events can be reported as soon as they occur. We model the problem as discovering dense clusters in highly dynamic graphs. Despite many recent advances in graph analysis, ours is the first technique to identify dense clusters in massive and highly dynamic graphs in real time. Given the characteristics of microblog streams, in order to find clusters without missing any events, we propose and exploit a novel graph property which we call *short-cycle property*. Our algorithms find these clusters efficiently in spite of rapid changes to the microblog streams. Further we present a novel ranking function to identify the important events. Besides proving the correctness of our algorithms we show their practical utility by evaluating them using real world microblog data. These demonstrate our technique's ability to discover, with high precision and recall, emerging events in high intensity data streams in real time. Many recent web applications create data which can be represented as massive dynamic graphs. Our technique can be easily extended to discover, in real time, interesting patterns in such graphs.

## 1. INTRODUCTION and MOTIVATION

Microblogging sites such as twitter.com have become a rich source of information about any "event", ranging from breaking news stories to earthquakes or information about local concerts. Empirical studies [10][11] show that (i) Twitter is often the first medium to break important events such as earthquakes, often in a matter of seconds after they occur and more importantly (ii) they highlight the need to discover all such events (and not just events related to earthquakes [10]) in real time from microblog streams. Note that by 'real time' we mean that events need to be discovered as early as possible after they start unraveling in the microblog stream. Such information about emerging events can be immensely valuable if it is discovered *timely* and made available.

[*] This work is done as part of the PhD at IIT-Bombay, India.

One obvious way to find information on microblogging sites is to use keyword search. There are many microblog search engines which allow users to find real-time microblogs relevant to a keyword query (e.g., twitter search). These search engines allow users to register their (continuous) keyword queries and return a stream of events, trends or news items relevant to the query. However, these search techniques do not help the user to "discover" the event but can be used to gather follow up information about the event. One could argue that the event could have been discovered by a continuous query with a keyword, say, "earthquake". However, note that a user would have to register a large number of such keyword queries to discover all possible types of events, something that is clearly not feasible.

The major challenge in achieving the goal of building a real time event discovery and tracking system lies in correlating the right microblog messages, among the hundreds of thousands of messages that are continuously being generated. The problem is exacerbated by the fact that the keywords used to describe the event might vary from one user to another and could also change over time due to the evolving nature of real time events. Hence classifier [10] or keyword search techniques may not be practical. This paper addresses these problems and presents a technique for discovering events in a microblog stream in real time.

Whenever an event happens, there will be a few keywords which will show burstiness (display a sudden jump in frequency). Hence a simple and obvious way to discover events is to keep track of the most popular words, something that is already done by twitter, and displayed as trending topics. A keyword (or a pair of consecutively occurring words) is recognized as a trending topic by Twitter if it is popular over a period of time. However, as reported in whatthetrend.com, several thousand tweets over a relatively short period of time are needed to identify an event as trending topic. Therefore, (1) using keywords appearing in 'trending topics' does not serve the purpose of discovering events in 'real time' (as by then the event would no longer be an emerging event) and (2) it is not necessary that all important events do become trending topics. Further, once a set of keywords becomes popular, they would remain so for a long time thereby overshadowing any new emerging events. Moreover, rather than reporting individual keywords or a pair of consecutive keywords it might be more meaningful and insightful to identify a set of correlated keywords (not necessarily occurring consecutively).

In order to identify an emerging topic, we need to identify a set of keywords which are *temporally correlated*, i.e., they show burstiness at the same time and are *spatially correlated,* i.e., they co-occur in temporally correlated messages from the same user. In order to capture these characteristics we use a dynamic graph model which uses the moving window paradigm and is constructed using the most recent messages present in the message stream. An edge between two nodes -- representing two keywords -- indicates that messages from a user within the recent sliding window involve the respective keywords. We use these properties to formulate our problem as that of cluster discovery in a dynamic graph. Figure 1 shows a partial graph induced by 6 real twitter messages (comprising 12 keywords). 6 of these keywords show burstiness (e.g., at least 2 occurrences). Keywords co-occurring together in messages from a user (within 6 messages) share an edge. We discover the cluster "earthquake struck eastern Turkey" in the graph, denoting an event. Two other keywords ("massive" and "moderate") were also bursty within the graph but they are not part of cluster (due to weak spatial correlation). When the window is moved at a fixed rate (oldest 2 messages expire and 2 most recent messages are added), a new keyword ("5.9") gets added to the cluster (denoting the intensity of earthquake).

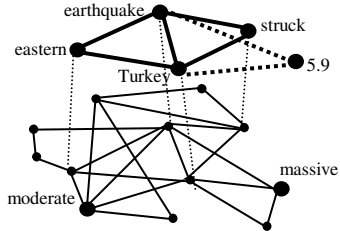

**Figure 1**: **Event cluster embedded in a graph drawn on Twitter data**

The example above highlights several issues central to our problem, specifically a) the cluster definition should be able to capture the imperfect correlation among keywords belonging to an event. Not all keywords are used by all the users (there is no edge between 'eastern' and 'struck'); b) we should be able to capture the evolving nature of events in a highly dynamic environment ('5.9' joined the cluster later); and, c) the identified events should truly be categorizable as emerging real-world events by filtering out spurious and unimportant events.

The graph is highly dynamic and complex, i.e., keywords (and associated edges) present in the graph get added and deleted at a fast pace; (Twitter reports more than 2300 tweets/sec [12] with potentially multiple simultaneous events present); the technique to identify the emerging events needs to be highly scalable. Specifically, it should be able to identify and maintain the events in a massive and highly dynamic graph.

## 1.1 Problem Formulation, Solution Ingredients

Let $S_i$ represent a set of keywords (potentially spread over multiple messages) from a unique user $i$ in a time window that spans from time $(t - \delta.w)$ to current time t, where $\delta$ represents unit time called *quantum* and $\delta.w$ is the length of the time window. Let $S_w^t = \{S_1...S_m\}$ be a set of keywords sent by $m$ unique users in the microblog stream in a given time window. As we are interested in discovering emerging events, $S_w^t$ contains the messages from a sliding window (of size $\delta.w$) over the message stream. Time unit $\delta$ denotes the fixed rate at which the window is moved.

*Correlated Keyword Graph* (*CKG*) *captures the properties of microblog contents.* We represent all the keywords, after removing stop words, appearing in the messages in the current window as nodes in an undirected graph, CKG (we use the terms *node* and *keyword* interchangeably in this paper). CKG *is a dynamic graph whose state at time t, is* $G^t = (V^t, E^t)$ *where* $V^t$ *is the subset of keywords appearing in message set* $S_w^t$. Thus, two keywords are said to be temporally correlated *iff* they appear in $V^t$ and are said to be spatially correlated if they have an edge between them in $E^t$. An edge links two keywords *iff* they both appear in a keyword set $S_i$ belonging to a user *i*.

Thus using the sliding window paradigm, a keyword is present in CKG if the keyword appears in at least one message in the current window. Since the window moves forward with time, CKG is highly dynamic where nodes and edges appear and disappear in real time. Further, a node in CKG can be either in a "high" state or a "low" state. A node moves into high state if there is a sudden increase in the frequency of its occurrence in the message stream.

Each edge in CKG is associated with a weight which signifies the probability of the words associated with the edge appearing in temporally correlated messages from a set of users. One of the challenges in working with highly dynamic microblog data is the size of the generated CKG. We overcome this challenge by constructing a much smaller Active CKG (AKG) from the original CKG such that (1) the clusters discovered in AKG are no different from those discovered in the CKG and (2) it is orders of magnitude smaller than the original CKG.

*Emerging events are identified by discovering clusters in CKG.* Given CKG, our problem of discovering emerging events can be mapped to identifying significant properties of the graph. For example, the burstiness of the keywords is captured by the state associated with a node. Temporal correlation can be captured by the moving window and spatial correlation can be identified by the weight associated with the edges. Using these properties, at a high level, our problem of event identification is similar to discovering a "cluster" within CKG. The cluster would consist of a set of keywords (e.g., "earthquake", "struck", "Turkey") where each keyword would be bursty and would exhibit temporal and spatial correlation with the other words in the cluster.

CKG is an undirected graph, i.e., is a tree of its biconnected components. A graph is said to be biconnected if for any pair of nodes in the graph there are at least two independent paths between them. Two paths are independent if they do not have any nodes in common except the end points. In a connected graph, two biconnected clusters can be connected with each other with just one path (had there been more paths between two clusters, they will merge into one cluster). We assume that nodes within biconnected components are more likely to be associated with the same event compared to nodes across components.

Biconnected components are the most encompassing forms of clusters in an undirected graph, next only to a connected graph itself being considered to be a cluster. However, if we choose to consider all biconnected components as our clusters, we may end up discovering massive and more often meaningless clusters in a large and dynamic graph. The other option is to consider only complete cliques, wherein each node is connected with all the other nodes in the clique, as clusters of interest to us. Complete cliques are more likely to represent interesting real world events but considering only complete cliques as clusters does not suit our scenario because a) different users may use different sets of keywords to describe the same event and b) keywords associated with an event change rapidly in the microblogging stream due to the evolving nature of real time events.

*Considering ½-quasi cliques (MQCs) as clusters of interest contributes to good precision and recall of discovered events*. As noted above, identifying events from the biconnected components in a CKG is likely to result in high recall (i.e., identify more real world events) but low precision (i.e., identify many non events as real-world events); the opposite is likely to be true for complete cliques. Therefore instead of finding either complete cliques or just biconnected components, we focus on ½-quasi cliques as our clusters of interest. A cluster is a $\sigma$-quasi clique if each node in the cluster is adjacent to at least $\sigma.(N-1)$ nodes in the cluster where $\sigma$ is a number between 0 and 1 and $N$ is the cluster size. When $\sigma$ is 1, the cluster is a complete clique. A biconnected component has $\sigma=2/N-1$. For a connected graph, the minimum value of $\sigma$ can be $1/N-1$. As explained above, none of the two extreme values of $\sigma$ is suitable in our environment. Therefore a natural choice is to set $\sigma$ to their mean in order to balance precision and recall. Hence, in order to discover meaningful clusters in a dynamic environment, we identify those components of a graph as clusters that have $\sigma >$

*(1/2+1/N-1)* or σ ≥1/2. We call these cliques *majority quasi cliques (MQCs)* since each node of the cluster is connected with a majority of the remaining nodes in the cluster.

***Exploiting* short cycle property (SCP) of MQCs makes event discovery a tractable and local problem.** It has been shown [14] that discovering ½-quasi cliques is an NP-complete problem even for static graphs. Fortunately, we are able to show that ½-quasi cliques possess a unique property which we call *short cycle property (SCP):* any edge in the cluster has at least one cycle of length at most 4 within the cluster. (In Section 4, we define the *short-cycle property* formally and show that (1) *SCP* is a necessary but not sufficient condition for *MQC*, (2) *SCP* is a sufficient but not necessary condition for bi-connected components, and, (3) *SCP* can be exploited to identify events by discovering clusters which possess the short cycle property (called *approximate MQCs* (*aMQCs*)).

The key advantage of using *SCP* for defining clusters is that we can discover dense clusters (*aMQCs*) efficiently and *locally* without using any global state information. For a dynamic graph, a cluster is said to be ***locally*** processable if for each incoming or departing node (or edge) to the graph, the cluster can be discovered by processing only its adjacent edges and nodes. Since these computations are *local* in nature, they are efficient, a pre-condition for discovering clusters in a highly dynamic graph. Further, multiple independent additions and deletions are allowed simultaneously on the graph. On the other hand, any processing which needs the graph to be stable (i.e., no addition or deletion is allowed in the graph during the course of computation) is called ***global*** processing. We believe that ours is the very first attempt to develop a technique to discover dense clusters in a highly dynamic graph. We propose efficient algorithms for discovering and maintaining the clusters in a dynamic graph as nodes and edges get added and deleted due to the moving window. We prove the correctness of our algorithms and experimentally show that our use of *aMQC* to define clusters helps us to discover emerging events correctly and efficiently.

***Globally consistent ranking of events can be achieved by exploiting local properties of clusters***. In order to consume events, a ranking function is needed such that important events are ranked higher compared to spurious or less important news. However, due to the highly dynamic environment and real time considerations, no ranking function which needs any global information can be used. We present a novel and highly efficient ranking function that ranks events by just exploiting the local cluster properties corresponding to each event, yet delivers a globally consistent ranking in a best effort manner.

Suppose two clusters discovered by us pertain to the same event but they could not get merged into a single event because (1) users used synonymous keywords to describe the event; (2) users indeed used different keywords, providing different perspectives about the same event; (3) the messages are posted in different languages. All these cases can be addressed by pre-processing the messages or post-processing the discovered clusters. For instance, one can use dictionary/thesaurus to address issues (1) and (3). For (2), clusters pointing to the same event should show temporal correlation. Therefore, one can post-process the discovered clusters (within a given time window) to correlate such clusters.

Further, suppose there is an ongoing discussion among tweeters about a controversial topic (resulting in many messages) but it is not a real world event. Typically, such "events" are ranked low compared to real world events due to their slow rate of spread. We may want to report even those events if they are ranked sufficiently high, but often one may want to ignore such events by post-processing the discovered clusters to identify such events. Post/pre-processing of keywords and discovered clusters and event categorization are orthogonal to the technique presented in this paper. It can be used to further enhance our technique and is part of our future work.

## 1.2 Research Contributions
- We present a new technique to discover and maintain dense clusters in massive and highly dynamic graphs in real time. In contrast, other clustering techniques, such as those based on data mining, are not only inherently slow in such environments they are also not suitable (details in Section 2).
- In Section 3 we present our strategy to construct a much smaller Active CKG (AKG) from the original CKG to help us efficiently discover and maintain the clusters, which is imperative in a dynamic environment.
- We model the problem of discovering the emerging events in real time in microblog streams as discovering approximate ½-quasi cliques, which possess the *short-cycle property*. This property is especially useful in highly dynamic microblog environments as it helps us maintain the clusters *locally* without using any *global* state information. We also prove the correctness of the algorithms (Section 4).
- We propose efficient algorithms for maintaining the clusters locally even under numerous additions and deletions of nodes and edges (Section 5).
- In Section 6, we present our ranking function such that more important events are highly likely to be ranked higher by just using local cluster properties.
- Through an experimental study of our technique using real twitter data, we demonstrate its ability to (1) discover the emerging events in real time -- with high precision and recall; (2) process at almost double the rate of current Twitter intensity on a machine of moderate configuration; (3) discover emerging events around the same time or much before it is seen on Google headlines; (4) discover additional events which do not appear in Google headlines (Section 7).

Discovering dense clusters in highly dynamic graphs efficiently and in real time has many applications in social networks, IP networks, telecommunication networks and for real time business analytics. Extant algorithms to discover dense clusters in dynamic graphs work on snapshot based techniques [2] and have severe limitations with regard to real time analytics. Our technique to discover clusters in massive and highly dynamic graphs in real time improves upon the state-of-the-art and can be easily extended for many such applications.

## 2. RELATED WORK
The notion of time evolving graphs where some communities show a burst in their behavior at certain points in time was first developed in [1]. This work, done in the context of blogs, developed techniques to study the evolution of connected component structures in time evolving graphs. A technique to find proximity between two nodes in time evolving bipartite graphs is proposed in [3].

In [2], authors propose a technique to discover keyword clusters in blogs [2] to identify a topic. The key difference between [2] and our work is that we discover keyword clusters in microblog streams which are very dynamic and under stringent real time constraints. The technique presented in [2] requires global computation of clusters in the graph where graphs are updated not in real time but on a daily basis. Similarly, there is a large body of work on analyzing the structure of communities and their evolutions in social networks [4][5]. These communities comprise humans (and not keywords), hence the real time constraint in this body of work is at a totally different scale compared to our problem setting. Hence our problem warrants a new approach.

Recent work on identifying *emerging topics* on Twitter data [13][17][18] has a problem statement similar to ours. These techniques use 'bursty keywords' in recent set of messages as *seeds* to identify emerging events. In [18] the authors describe a series of heuristics to identify 'emerging terms'. The technique is computationally expensive for real time analysis due to the iterative method that is employed to compute an 'authority score'. Further, the concept of user's authority to identify emerging terms may not be applicable in most real world situations. [13] reports a *pair of keywords* (based on correlation) as an emerging topic. At least one of these two keywords should be among the 'bursty keywords'. [17] reports a cluster of keywords (at least one of them has to be 'bursty'). However, both the techniques suffer from multiple limitations; 1) the performance is highly sensitive to the value of keyword burstiness threshold; 2) in the presence of multiple events, identifying co-occurring disjoint subsets from all the bursty keywords [17] or identifying co-occurring pairs based on time series analysis [13] are computationally expensive techniques. Events are not ranked in [17] therefore making the consumption of events untenable in the presence of multiple events. Further, the methods in [13][17][18] are not able to capture the evolution of an event as all these techniques use *seed* keywords to identify events. These techniques are essentially based on post-hoc analysis but they highlight the importance of identifying keywords (nodes) and their correlation (edges among keywords) as the basis for identifying an emerging topic.

## 3. REDUCING GRAPH SIZE: CKG to AKG

Due to the high rate of arrival of messages in microblogs, the CKG generated from the microblog stream can quickly become very large. Hence we first generate a manageable sub-graph, AKG, from the original CKG so that our cluster discovery problem becomes tractable.

### 3.1 Identifying AKG Nodes

We use a 'hysteresis' based approach. Let the CKG be denoted by $G^o$. Each node in this graph represents a keyword in the data stream (after removal of stop words). As we are interested in finding the emerging events, a natural way is to pick only *active* keywords in $G^o$ which show an upward trend in their burstiness, i.e., frequency of their occurrence across different messages during a quantum, crosses a given threshold. Towards that end, we construct a subgraph, called AKG using only the active keywords and ignoring all the other keywords and their associated edges. Let $G$ be the AKG induced by $G^o$ after removing the non-bursty keywords. Notice that threshold in our case is set to identify keywords that need to be *excluded* from $G$ and hence it is significantly low. However, given a properly set threshold, $G$ will still be significantly smaller in size as compared to $G^o$, since only a small number of keywords would show burstiness. Because the burstiness threshold is low, the graph $G$ contains all the keywords associated with an *emerging* event. We can subsequently use $G$ to identify the events without impacting precision and recall.

In order to identify bursty keywords, we use a two-state automaton where each keyword is either in a *low* state or a *high* state. A keyword moves from *low* state to *high* state (i.e., added in AKG) if during a quantum it shows burstiness, i.e., it appears in more than $\gamma$ different users' messages. We call $\gamma$ the *high state threshold (HST)*. All other keywords are in *low* state. A keyword in *high state* may remain bursty or may become non-bursty in subsequent quanta. As we are interested in finding emerging events, we are specifically interested in finding keywords which are moving from *low* to *high* state.

In order to discover meaningful clusters in $G$, we need relative stability in the graph. Hence, a keyword which has moved to AKG remains in AKG as long as it is part of an event cluster irrespective of its frequency of occurrence in subsequent quanta However, as we maintain the graph over a sliding window, we remove all the stale keywords, i.e., those keywords which have not occurred in any of the last $w$ quanta, from AKG.

For the keywords present in AKG, we update their status (i.e., remove them from AKG) using a *lazy update* principle, if needed, for only those nodes which are (1) in AKG and also occur in the messages present in the current quantum and (2) nodes adjacent to nodes identified in (1), as their correlation can change. One can see that in a given quantum only these nodes can be removed from a cluster (due to change in correlation). A departing keyword from a cluster is removed from the AKG if it is not part of any other event cluster. Notice that, as we explain in Section 3.2, a keyword which is not part of any cluster cannot become part of another cluster unless it exhibits a high frequency behavior. At that point, the keyword is moved into AKG anyway.

The above technique helps us to smooth the movement of keywords from *high* to *low* state or vice-versa and is more efficient and scalable compared to the time series analysis as required in [13]. We can compute the state of each arriving keyword at the end of the quantum in $O(1)$ time. Once the nodes in the sub-graph have been identified, the next step is to find the edges between these nodes.

### 3.2 Identifying AKG Edges

The guiding principle for creating an edge between two nodes in the sub-graph $G$ is that*, in the current time window, messages from a significant number of users should have both the keywords*. Therefore, we associate a correlation measure with the edge connecting the two nodes and place an edge between the two keywords (present in AKG) if the correlation between them is above a threshold. The correlation is computed by associating a set of user ids with each keyword. This set $U_1$ (called the *id set*) associated with a keyword $n_1$, contains the ids of all those users who used this word in the current window. Given sets $U_1$ and $U_2$ for a pair of nodes $n_1$ and $n_2$, we can find their correlation by using the *Jaccard coefficient,* which is defined as the size of the intersection divided by the size of the union of the two sets: $|U_1 \cap U_2|/|U_1 \cup U_2|$. We call it *edge correlation* (*EC*). Notice that a high value of the *EC* would imply that the two keywords have been used together by a large proportion of users and would hence imply a strong correlation. We use user ids as opposed to message ids so as to avoid the case of a single user flooding the same message multiple times leading to high correlation between nodes of the message. However, if we use user id, the strict message based spatiality is not valid (it is not necessary for a user to mention all the keywords in the same message). Hence, spatial correlation is not confined to a message but to a user and keywords from a user may be spread over multiple messages albeit within a given quantum of size $\delta$.

Since AKG contains all the keywords in the *high state*, it would be costly to compute the correlation of all pairs of nodes in AKG. Hence, we next address the following challenges: (1) Identify those pairs of nodes whose correlation is likely to be above the threshold and; (2) Find the correlation of the selected nodes.

### 3.2.1 Identifying node pairs for EC Computation
As mentioned earlier, each keyword is associated with an *id set*. For keywords appearing in the *last* quantum, we construct two sets with the aid of *id set;* (1) Keywords that are in the high state (the size of the associated *id set* is $\geq \gamma$) and (2) keywords that were already in AKG and have also appeared in at least one of the messages that arrived in the last quantum. Note that a keyword may appear in both the sets. For all the keywords in set (1), we compute the correlation only among them. If the *EC threshold* is $\lambda$ and if the *correlation* between two keywords is above $\lambda$ we place an edge between them. It is intuitive to see that *new* keywords, entering into AKG do not have temporal and spatial correlation with any other keyword present in the AKG except those in set (1). For all the keywords identified in set (2), we update their correlation with their neighbors. Any other pairs of keywords would not have their correlation changed.

Thus, using this mechanism we drastically reduce the number of *EC* computations that we need to do at the end of each quantum. As described next, we use the Min-Hashing scheme [6] to compute the *EC* efficiently.

### 3.2.2 Efficient computation of EC
We assign a hash value to each unique user in a quantum. Assuming that the number of unique users per quantum is no more than $2^n$, we choose the hash value for each message independently and uniformly from a range $(0, 2^{2n})$ so as to avoid the birthday paradox (hash collision) [8]. For each keyword, we then keep track of the minimum hash value (*Min-Hash*) among all the user ids present in its *id set*. Now, for each pair of nodes $n_1$ and $n_2$, the probability of both $n_1$ and $n_2$ having the same *Min-Hash* value is exactly equal to their *Jaccard similarity coefficient* [7] or *EC*. The reasoning is as follows: The *Min-hash* value will be the same if the *id* with minimum hash value is common to both the *id set* nodes, i.e., it belongs to the set $|U_1 \cap U_2|$. Since the total size of both the sets is $|U_1 \cup U_2|$, the probability of having the same *min-hash* value is $|U_1 \cap U_2|/|U_1 \cup U_2|$. However, in order to avoid false negatives, instead of keeping track of only a single *Min-Hash* value for a node, we keep track of $p$ *Min-Hash* values (i.e., the $p$ minimum hash values amongst all the user ids in the union of *id set*). We add an edge between two keywords in $G$ if there is at least one common entry in their $p$ *Min-Hash* values. The value of $p$ depends on the *EC* threshold $\lambda$ and *high state* threshold $\gamma$; for a uniform distribution, the expected number of trials before a match occurs is $1/p.\lambda$. Value of $p$ is set to $min(\gamma/2, 1/\lambda)$. Due to this mechanism, we can compute the correlation between two nodes in an efficient manner with a very small probability of false negatives and false positives [7].

In summary, we first significantly reduce the number of pairs of nodes whose correlation needs to be computed and then for the identified pairs we find their *Jaccard coefficient* efficiently in $O(p.log(p))$ time where $p$ is a constant.

Thus, the tunable parameters and thresholds affecting the AKG are $\gamma$, $\lambda$, $\delta$ and $w^* \delta$. One can argue that it is imperative to set the thresholds ($\lambda$ and $\gamma$) correctly to include an edge and a node in the AKG. The discovery of an event ultimately depends on what nodes and edges are present in the graph, which in turn depends on these threshold values. For timely discovery of events these thresholds are kept low and they are just the qualifying thresholds for any edge or node to be in the AKG. If the $\gamma$ is high, only very popular keywords reach the high state. It compromises our ability to identify the emerging events in a timely manner. Further, since not all keywords are used by all the users, the threshold for each individual keyword has to be low. For the same reason, $\lambda$ has to be relatively low. Therefore, thresholds are set such that they just filter out completely unwarranted nodes and edges and not tuned such that nodes and edges left in the graph automatically result in events. However, with low threshold, many more keywords move into high state. Therefore, the events are identified by discovering a particular class of clusters (aMQCs) as explained in Section 4.

## 4. CLUSTER DISCOVERY
Once the graph is in place, we can use many standard cluster finding algorithms [2] to find a cluster of keywords corresponding to an emerging event. However, approximation algorithms for finding dense clusters in a graph operate on the entire graph (i.e., graph needs to be stable during the computation) and are not efficient [2]. We hence propose the novel *short cycle property* (SCP) in Section 4.1 which helps us discover dense clusters (i.e., aMQC cliques in our case) efficiently and in real time.

In Figure 1, a cluster with 4 keywords ("earthquake", "struck", "eastern", "turkey") exists at time *t*. At time $t+\delta$, we could update the cluster *with* keyword "5.9" since it was forming a cycle of length 3 with nodes ("earthquake", "turkey"). If the edge between "earthquake" and "turkey" would not have existed, even then keyword "5.9" would have joined the cluster due to the formation of cycle of length 4, via keywords "eastern" or "struck". Hence, due to the existence of a *short cycle* within the nodes of the cluster, we could update the cluster without re-computing it on the entire graph. As the graph changes, *SCP* ensures that only incremental computations are performed for those nodes and edges which need to be updated, while simultaneously ensuring the correctness of result. We provide the analysis and correctness of our approach in Section 4.2 and Section 4.3 respectively.

### 4.1 Short-cycle property
A graph is said to possess the short-cycle property if for any two adjacent nodes $n_1$ and $n_2$ in the graph, in addition to the direct edge between $n_1$ and $n_2$, there exists at least one more path of length at most 3 between $n_1$ and $n_2$, i.e., $n_1$ and $n_2$ are part of a cycle of length at most 4. More formally, *short-cycle property* in a cluster $C(V,E)$ in graph $G$ is defined as follows: if $\{v_i, v_j\} \in V(C)$ and $(v_i, v_j) \in E(C)$ then $\exists (v_i \to v_j)$ s.t., $1 \triangleleft v_i \to v_j \leq 3$.

**Definition 1:** The diameter of a graph $G(V, E)$ is defined as $D(G) = \max_{u,v \in V(G)}\{d(u,v)\}$ where $d(u,v)$ is the distance between any two nodes $u, v$, belonging to the graph. The diameter of a complete clique is 1.

Theorem 1 states that *SCP* is a necessary property for *MQC*s.

**Theorem 1:** For a majority quasi clique $G(V,E)$ with $\sigma \geq \frac{1}{2}$, $\forall v \in V(G)$, $v$ participates in a cycle of length at most 4.

**Proof:** Let us denote the neighbor set of node *u* as $A(u)$. $u \in G_{MQC} \Rightarrow |A(u)| \trianglerighteq \lceil (N-1)/2 \rceil$ where $|V(G)| = N$. For a graph $G(V,E)$ with $\sigma \geq \frac{1}{2}$, $D(G)=2$ [15]. Hence $\forall_{u,v \in V(G)}\{d(u,v)\} \leq 2$.

**Case1:** $d(u,v)=2$;

$u, v \in V(G_{MQC}) \Rightarrow \exists n_0 \mid (u, n_0) \& (n_0, v) \in E(G_{MQC})$. We claim that for pair of nodes $\{u,v\}$, there is at least one more common neighbor apart from node $n_0$.

Let us define $S_{i \in \{u,v\}} = A(i) - n_0; \{i, n_0\} \notin S_i \Rightarrow |S_u \cup S_v| \leq N-3$ (1).

Eqn (1) holds since nodes $u$, $v$ and $n_0$ are not part of $|S_u \cup S_v|$.

Since $|V(G_{MQC})| = N$ and, $|S_{i \in \{u,v\}}| \geq \frac{1}{2} \lceil N-1 \rceil - 1 \Rightarrow |S_u \cap S_v| \geq 1$, otherwise Eqn (1) will not hold. In other words, $u$ and $v$ have at

least one more neighbor apart from node $n_0$. Hence, there exists a cycle of length 4 between any pair of nodes $\{u,v\} \in V(G_{MQC})$.

**Case 2:** $d(u,v) = 1$;

$u,v \in V(G_{MQC})$ i.e. u and v have an edge between themselves. In this case, without loss of generality, for any neighbor $n_0$ of node $u$, $d(n_0,v) \leq 2$. Hence $\{u,v\} \in V(G_{MQC})$ are part of cycle of length 4.

Hence in both these cases, node $v$ ($u$) is part of a cycle of length at most 4 or, in other words, for any $(u, v) \in E(G_{MQC})$, there exists another path between them of length at most 3 within the cluster.

The *short cycle property (SCP)* of MQCs, as we explain next, radically simplifies the cluster discovery problem. Capitalizing on the *SCP* we can add a new node to the existing clusters *locally* (i.e., by just processing the edges adjacent to it) as follows. For each new keyword *n* that is moving into *high* state, if it shows a correlation with $n_1, n_2,\ldots n_k$ ($k > 1$) keywords (nodes) in graph G, we check if each node pair $n_i, n_j$ ($1 \leq i, j \leq k$):

*Rule R1*: Has at least one more common neighbor OR
*Rule R2*: Has an edge between them.

In either case, we add the new node to the cluster that both these nodes are already part of. If these two nodes ($n_i, n_j$) are not part of any cluster, we initialize a cluster with four nodes if it satisfies (*R1*) or three nodes if it satisfies (*R2*). For *k=2*, as shown in Figure 2, an incoming node *n*, forms a cluster (a) as *n1 and n2* have a common neighbor $n_c$ (*R1*) or cluster (b) as $n_1$ and $n_2$ have an edge between them (*R2*). If the incoming node shows correlation with zero or one node, we simply add that node (and edge) in *G* and do nothing.

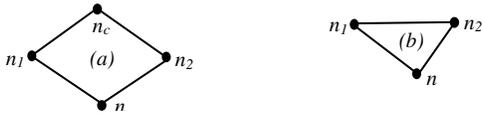

**Figure 2: Clusters formed due to short-cycle property**

We check *R1* and *R2*, for each departing node (node which moves from *high* state to *low* state), where existing clusters can be either re-clustered into smaller clusters or dissolved (Section 5.3). For all edges adjacent to the arriving or departing node, we consider two adjacent edges at a time (total O ($k^2$) pairs of edges if there are *k* adjacent edges to that node). We check if nodes which these two edges are adjacent to, satisfy either *R1* or *R2*.

Therefore without processing any other nodes and edges in the graph, except the pairs of edges adjacent to the node under consideration, we can discover a cluster that satisfies *SCP* and thus an *aMQC*. Thus, due to this special property, we can discover the approximate 1/2-quasi cliques in the dynamic graph AKG by performing just local computations. At each time quantum, we do a total of $O(k^2NC)$ computations where *N* is the total number of nodes changing their status (to *high* or *low*), *k* is the average number of edges adjacent to these nodes and C is the average cluster size a node (among N nodes) is participating in. Now, by our definition, both *k* and *N* are fairly small compared to the number of keywords present in the message stream in a given time window. Further, as shown by our experiments, the average cluster size is very small compared to the size of the graph (less than 7 keywords/cluster).

For MQC, short cycle property is a necessary but not sufficient condition. For the cluster in Figure 3(b) (including new edges), each edge participates in a cycle of length 4 within the cluster but the cluster is not MQC. If we identify the cliques based on the short-cycle property, while we will not miss any MQC, we may collect some extra clusters which are not MQCs. As shown in [14], discovering MQC is NP-hard even for static graphs. Therefore, discovering clusters based on SCP discovers the MQCs not only with a good approximation bound, but also very fast and we can discover dense clusters with just *local* computation.

## 4.2 Analysis of Approximate MQCs

As we proved in Section 4.1, SCP is a necessary condition for *MQCs*. Once an aMQC is discovered based on SCP, one can efficiently identify if it is MQC in $O(N^2)$ time where N is the number of nodes in the discovered cluster. We check if each node belonging to the cluster has edges with at least *(N-1)/2* nodes in the cluster. However, with dynamic graphs, we face challenges which are different from stable graphs as depicted below.

**Example 1:** Let us consider a MQC of size 7 as in Fig 3(a) which is reported as a cluster. Since the clique size is 7, each node has to be connected with at least $\lceil 6/2 \rceil = 3$ of the nodes in the clique. Now, if a $8^{th}$ node joins the clique (due to the existence of *short cycle* with nodes in the cluster), for the original cluster to be continued to be considered MQC, each node should have connection with $\lceil 7/2 \rceil = 4$ nodes. Hence, the new node should a) have edges to at least 6 of the existing nodes in the cluster or b) have connection with any of the 4 nodes in the cluster along with at least 1 more *new* edge among already existing nodes in the cluster. Point (b) not only makes the computation of MQC exponential, it is also an unnecessary requirement since the cluster with existing 7 nodes is already reported. On the other hand, the requirement to have an edge with almost all the other existing nodes is too stringent for admitting any node in the cluster as the keywords belonging to an event may keep on changing. This example shows that since real time events evolve continuously, using MQC as our cluster definition restricts our capability to capture dynamic events.

**Example 2:** We show two separate clusters (MQC clusters) in Figure 3(b), both discovered based on *SCP*. Now assume that two new edges emerge among two clusters, as shown in Figure 3(b), forming a short cycle between the nodes belonging to separate clusters therefore, due to *SCP,* merging these two clusters into one. Now a) either we stop reporting both of these earlier clusters as events since the merged cluster is no longer an MQC or b) we keep on reporting earlier clusters as separate clusters.

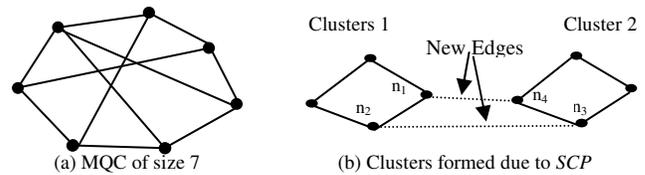

(a) MQC of size 7      (b) Clusters formed due to *SCP*
**Figure 3: Clusters discovered based on *SCP***

Both of these scenarios point out the following issues: In case of (a) we stop reporting the events already reported, the basis for which is still intact. The nodes in the event continue to show correlation with the same set of nodes as in the erstwhile clusters (one may, however, stop reporting the event if any node/edge disappears); In case of (b) maintaining such distinction will not only be computationally expensive in a dynamic environment (we need to identify all sub-cliques in a discovered cluster such that these sub-cliques are MQCs), it will be erroneous also (nodes $n_1$, $n_2$, $n_3$, $n_4$ would be reported as a separate cluster). On the other hand, emergence of new edges among the nodes belonging to two events, both discovered close to each other in real world time, points to a strong temporal and spatial correlation.

However, if we relax the requirement of having MQC and instead consider *aMQC* based on *SCP*, as our clusters of interest, we not

985

only capture the evolving nature of real time events in a fast moving environment, we discover the clusters more efficiently. *aMQC* cliques allow incremental evolution of clusters therefore capturing the evolving nature of the real time events. Hence, in Example 1, a new node is able to join the cluster due to *SCP* indicating the continuous evolution of the real time events. Similarly, in Example 2, two clusters exhibiting strong temporal and spatial correlation are allowed to merge into one event.

However, if the evolved cluster is sparse, it is more likely to be ranked lower due to its inherent sparse nature. Our ranking function (Section 6) ensures the quality of discovered events by ranking more dense clusters higher. Hence, the *SCP* helps discovering dense clusters in a scalable and efficient manner.

Therefore, even though one can efficiently identify MQC from an aMQC, due to the dynamic nature of the graph and the evolving nature of the events, *SCP* is the only cluster property that we enforce while discovering clusters in a dynamic graph. The aMQCs based on *SCP* ensure that no MQC based clique is missed. At the same time, the clusters thus discovered are biconnected components as *SCP* is a sufficient (but not a necessary) condition for biconnected components as shown in Section 4.3. The bi-connected property of clusters is helpful in maintaining the events efficiently in a highly dynamic graph as explained in Section 5.

## 4.3 Correctness of our Approach
We next present the main properties of our clusters (aMQCs) and give a correctness proof for our approach. We first prove that clusters discovered by us are bi-connected. Thereafter we give a proof of correctness of our approach, i.e., the clusters discovered based on *local* processing of nodes and edges are unique and consistent with similar clusters discovered on a time invariant instance of the same graph.

***Theorem 2***: *If we discover the clusters based on SCP, the resulting clusters will be bi-connected.*
***Proof***: The proof is by induction and is based on *Lemma 1*.   □

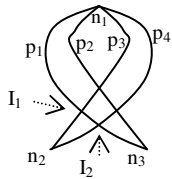

**Figure 4**: Independent path examples

***Lemma 1***: *Given any three nodes $n_1$, $n_2$ and $n_3$ belonging to a cluster, there exist two independent paths from $n_1$ to the other two, i.e., there exit two paths, one from $n_1$ to $n_2$ and another from $n_1$ to $n_3$ which are independent from each other.*

***Proof Sketch***: As the cluster is bi-connected, there exist two independent paths from $n_1$ to $n_2$ and from $n_1$ to $n_3$. As shown in Figure 4, let the two paths from $n_1$ to $n_2$ be named $p_1$ and $p_2$ and those from $n_1$ to $n_3$ be named $p_3$ and $p_4$. $p_1$ and $p_2$ are independent paths and hence do not intersect with each other. The same holds for paths $p_3$ and $p_4$. Now, there are three cases:
C1: None of these 4 paths ($p_1, p_2, p_3, p_4$) intersect with each other. Hence there exist 2 independent paths from $n_1$-$n_2$ and $n_1$-$n_3$.
C2: Only one pair of paths intersects each other. Without loss of generality, say paths $p_1$ and $p_3$ intersect with each other. Hence, there exist 2 independent paths from $n_1$-$n_2$ ($p_2$) and $n_1$-$n_3$ ($p_4$).
C3: Both of these pairs of paths intersect with each other. Therefore, there must exist at least 2 intersection points. Let's call them $I_1$ and $I_2$. We can always construct two independent paths from $n_1$ to $n_2$ and $n_1$ to $n_3$ as follows: $n_1$-$I_1$-$n_2$ and $n_1$-$I_2$-$n_3$. Independent paths can be constructed even if $I_1$ and/or $I_2$ themselves are a sequence of nodes by extending the same argument. The detailed proof is omitted in the interest of space.

**Correctness of Local Computation**: We now prove that our cluster computation is correct, unique and consistent.

***Lemma 2:*** *The locally discovered clusters are consistent with any global computation of clusters on the same graph.*

***Proof***: The proof is by induction.   □

As we see above, node *n* need not be present in the graph at the time of computation of cluster *C*, and as and when it arrives, by just processing its adjacent edges, we update the cluster. Now, suppose, an incoming (or departing) node *n* is adjacent to nodes $n_1,..n_k$. $e_i$ is an edge from node *n* to node $n_i$

***Lemma 3:*** *Each pair of edges ($e_i, e_j$), $1 \leq i,j \leq k$, $i \neq j$ will merge at most two clusters (for incoming node).*

***Lemma 4:*** *Each pair of edges ($e_i, e_j$), $1 \leq i,j \leq k$, $i \neq j$ will break the cluster into at most two clusters(for departing node).*

However, it may be the case that one or more of the resulting sub clusters no longer remain aMQC as *SCP* may no longer hold for the cluster. The process to check this is described in Section 5.3.

***Lemma 5***: *For all pairs of edges ($e_i, e_j$), $1 \leq i,j \leq k$, $i \neq j$ adjacent to node n, the final cluster(s) do not depend on the order in which each of these pairs is considered.*

Similarly, for an incoming/departing edge *e*, adjacent to nodes $n_1$ and $n_2$, clusters are maintained by considering all pairs of edges ($e, e_i$) where $e_i$ ($\neq e$) is an edge adjacent to either node $n_1$ or $n_2$ (as outlined in Section 5). Therefore, Lemmas 3, 4 and 5 are applicable for edge addition/deletion as well.

***Theorem 3:*** *The locally discovered clusters result in the unique clustering for a given graph.*

***Proof***: Follows as a corollary of Lemmas 2, 3, 4 and 5.   □

In summary, the properties of a *cluster C* discovered based on SCP are:
*P1*: *C* is an a*MQC* as *SCP* is a necessary (but not sufficient) condition for MQC.
*P2: C* is a bi-connected cluster as *SCP* is sufficient (but not necessary) condition for bi-connected clusters.
*P3: C, discovered locally* with the aid of *SCP,* is consistent with global computation on the same graph, is correct and unique.

## 5. CLUSTER MAINTENANCE
We now present the details of the algorithms for node/edge addition/deletion. These operations do not require any global computation. We first prove a property of aMQCs below.

***Lemma 6***: *Two aMQCs which share an edge are merged to form a single aMQC*
***Proof Sketch***: Consider two aMQC clusters $C_1(V_1,E_1)$ and $C_2(V_2,E_2)$. Let edge $e_1$ between nodes $n_1$ and $n_2$ be common between $C_1$ and $C_2$, i.e., $e_1 \in E_1$ and $e_1 \in E_2$. If $C_1$ and $C_2$ are merged to form a single cluster $C(V,E)$ then the merged cluster will be an aMQC and satisfy all our cluster properties (*P1,P2 P3*). As explained next, we use this property to merge clusters as new nodes and edges are added to the graph.

## 5.1 Node Addition
The node addition algorithm is based on the *SCP*. Hence for a new node $n_1$ to be made a part of cluster $c_1$, it should have *edges* to at least two nodes $n_2$ and $n_3$ within the cluster. In order to satisfy the *SCP*, either (a) $n_3$ or $n_2$ should be neighbors of each other or (b) $n_2$ and $n_3$ should be connected by a path of length two. The node addition algorithm can be stated as follows:

**Algorithm**: NodeAddition
Let V' be the set of nodes which are incident on the newly added node $n_1$. For all pairs of nodes $(n_2, n_3) \in$ V',
  if $(n_2, n_3) \in$ E, form a new cluster from $n_1, n_2$ and $n_3$.
  Find all nodes N which are adjacent to both $n_2$ and $n_3$.
  Form a new cluster from the nodes $n_1, n_2, n_3$ and $\forall n_4 \in$ N.
Merge the clusters using the cluster merging algorithm till no more merging is possible. In Figure 5(b), when a new node *n* arrives, it has edges to node *1* and *2*. These two nodes have a common neighbor (node *4*). Hence, a new cluster (*1, 2, 4, n*) is formed due to presence of *SCP*. Since this new cluster shares an edge (*1, 4*) with $C_1$, it is merged with $C_1$. This merged cluster is again merged with $C_2$ due to edge (*2, 4*) resulting in cluster $C_4$. Being based on *SCP*, we can see that the newly formed clusters will satisfy *P1, P2 and P3*.

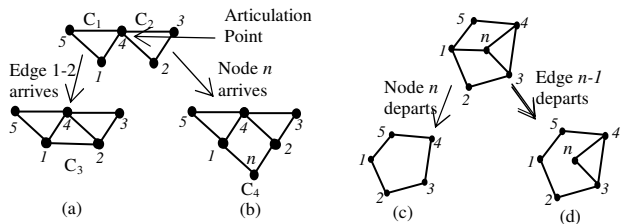

(a)  (b)  (c)  (d)

**Figure 5**: **Node/edge addition and deletion examples**

## 5.2 Edge Addition

The edge addition algorithm also tries to ensure that the fresh clusters formed due to the new edge satisfy the *short-cycle property*. We present the algorithm and then prove its correctness. Let a new edge $e_1$ ($n_1$, n2) be added to the graph. Notice that both the nodes $n_1$ and $n_2$ already existed in the graph G(V,E).

**Algorithm**: EdgeAddition
$\forall n_3 \in$ V | $(n_1, n_3) \in$ E do
  $\forall n_4 \in$ V | $(n_2, n_4) \in$ E do
    if $n_3 = n_4$ or $(n_3, n_4) \in$ E, form a cluster of $n_1, n_2, n_3, n_4$.
Merge the clusters using the cluster merging algorithm.

The EdgeAddition algorithm works in two phases. In the first phase it forms all those clusters which satisfy the *short-cycle property* with the newly added edge. Once it has formed these clusters, it merges them using the cluster merging algorithm. The clusters formed during the first phase satisfy the *short-cycle property* and hence satisfy all our cluster properties. As per *Lemma 6*, the clusters discovered during the second phase would also satisfy our cluster properties. Hence the edge addition algorithm discovers correct clusters.

In Figure 5(a), a new edge (*1,2*) arrives. In phase 1, we create three clusters namely (*1,2,4*), (*1,2,4,5*) and *(1,2,3,4,)*. In phase 2, these aMQCs are merged (*Lemma 6*) to form the cluster $C_3$.

## 5.3 Node Deletion

When a node is deleted from the graph all the incident edges on that node also get deleted. As a result of this the clusters in which the node participates could get split into one or more smaller clusters. Due to *short-cycle property,* standard depth first search based techniques to partition a biconnected component do not work in our environment. Thus the major task associated with the deletion of a node is to ensure that the correctness of the clusters is maintained post the deletion of the node. This implies that the partitioned clusters satisfy the *short-cycle property*.

Thus a cluster will not get dissolved/split if (1) each edge in the cluster is part of a *short-cycle* within the cluster and (2) if the cluster does not have an articulation point. Notice that after the deletion of a node, a cluster could satisfy (1) and still have an articulation point as shown by the example in Figure 6. In the figure, initially the graph consists of a single cluster consisting of all the nodes. When node 9 gets deleted, the cluster gets split into two as node 3 now becomes an articulation point (Cluster 1 – nodes 0,1,11,10,2,3 and Cluster 2 – nodes 4,5,8,7,6,3). Hence whenever a node gets deleted, we need to perform two checks:

*Cycle Check*: find the edges which do not participate in a *short-cycle,* i.e., a cycle of maximum length four; and

*Articulation Check*: find if any articulation points are generated in the cluster.

Before we explain the algorithm for dissolving clusters, we first present a property satisfied by the set of nodes that can become articulation points due to the deletion of a node and its incident edges. Once we identify this set we can restrict the articulation check to the nodes in this set thereby improving our efficiency.

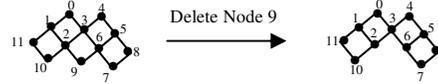

**Figure 6: Breaking of cluster due to node deletion**

*Lemma 7*: Let nodes $n_1, n_2, n$ and $n_c$ belong to cluster C (See Figure 2(a)). n has *only* two incident edges $e_1(n, n_1)$ and $e_2(n, n_2)$. $n_1$ and $n_2$ have a common neighbor $n_c$. Let the node n along with its edges $e_1$ and $e_2$ be deleted. No other node except $n_c$ can be the articulation point.

*Proof Sketch*: Let the articulation point in the cluster be a node $n_a \neq n_c$. Since $n_a$ is part of an aMQC cluster, it participates in a *short-cycle*. That *short-cycle* cannot have either $e_1$ or $e_2$ (otherwise node n would have had at least one more edge adjacent to it). Therefore, $n_a$ continues to be the part of a cycle and cannot be an articulation point. □

In case there is no node adjacent to both $n_2$ and $n_3$ but there exists a direct edge between $n_2$ and $n_3$ then it can be shown that both $n_2$ and $n_3$ will become the articulation points after the removal of $n_1$. This can be proved using arguments similar to those given above.

It is important to note that the node(s) suggested by *Lemma 7* will be an articulation point if there are no alternate paths between $n_1$ and $n_2$ except the one via $n_c$. In other words if there is a direct edge between $n_1$ and $n_2$ or if there are multiple nodes neighboring $n_1$ and $n_2$ then $n_c$ cannot be an articulation point. This is intuitive from the above and hence the proof is omitted.

The articulation check is performed for each pair of edges adjacent to node *n,* which participate in a *short-cycle.*
The node removal algorithm uses the *Lemma 7* to restrict the articulation check to a small set of nodes. We now explain the details of this algorithm. Let the graph G(V,E) consist of a cluster C having nodes V(C) and edges E(C).

**Algorithm**: NodeDeletion
Let $V_I \subseteq$ V s.t. $\forall n_2 \in V_I | (n_1, n_2) \in$ E(C)
Delete node $n_1$ and all its incident edges
**Cycle Check**
$\forall n_2 \in V_I$ do
  $\forall (n_2, n_3) \in$ E(C) check if the edge $(n_2, n_3)$ has a path of length

at most 3 within the cluster
If not remove edge from cluster and add $n_3$ to $V_I$
If yes, check if at least one edge of the cycle is shared with another cycle of the original cluster of length at most 3.
If not, create an independent cluster from this cycle.

**Articulation Check**:
$\forall\ n_2$ and $n_3 \in V_I$ do
 if $(n_2, n_3) \notin E(C)$ and there exists exactly one common neighbor of $n_2$ and $n_3$ then check if there is any path from $n_2$ to $n_3$
 if not, split C into two clusters – one consisting of node $n_2$ and nodes in V(C) reachable from $n_2$ except via $n_3$. The remaining nodes will be part of another cluster.
 if $(n_2, n_3) \in E(C)$ and there is no other path from $n_2$ to $n_3$ of length at most 3, then check if there is any path from $n_2$ to $n_3$
 if not, split C into two clusters – one consisting of node $n_2$ and nodes in V(C) reachable from $n_2$ except via $n_3$. The remaining nodes will be part of another cluster.

In Figure 5(c), node *n* is removed. Set $V_1$ contains nodes 1, 3 and 4 in the beginning. Node 2 and 5 are also added to set $V_1$ as described in *cycle-check*. Since none of the nodes participates in a *short-cycle,* the cluster is no longer an *aMQC* and is discarded. In Figure 6, when node 9 is deleted, it generates an articulation node (node 3), and gets split into two clusters as described above. Articulation check is done for smaller set of nodes selected based on *Lemma 7*. If we find an articulation point then we split the original cluster around the articulation point.

Thus the above algorithm helps to finds the new clusters locally by only focusing on the nodes taking part in the original cluster. Further, the algorithm needs to evaluate all the nodes of the original cluster if and only if we find some articulation point. Articulation points are used to discover bi-connected components in static graphs. We present algorithms such that we use articulation points to efficiently maintain the clusters *locally* as described above. Articulation points could be efficiently exploited due to the bi-connected property of aMQCs. In most of the cases the algorithm is able to discover the new clusters by visiting a fraction of the nodes of the original cluster.

## 5.4 Edge Deletion
The edge deletion algorithm is very similar to that of node deletion. When an edge $e_1(n_1, n_2)$ is deleted, we need to perform a cycle check to find all the cycles of length at most 4 that could have been broken due to the deletion of this edge. In Figure 5(d), edge (n, 1) is deleted. Set $V_1$ (in NodeDeletion algorithm) is initialized with nodes {1,n}. In *cycle-check* phase, a smaller cluster with nodes (3, 4, n) is created since nodes 1, 2 and 5 are no longer part of a *short-cycle.*

## 6. RANKING EMERGING EVENTS
We discover emerging events in real time in a microblog stream. It is important to rank these discovered events in order to present these events to users in a comprehensible manner such that relatively more important events are ranked higher. Further, due to overwhelming pace at which the messages are generated in a microblog stream, it is entirely possible that some spurious events may get discovered due to accidental formation of a cluster, for instance because of presence of some popular keywords in the graph. Hence, our goal is to not only identify *real* events but to rank relatively more important events higher.

We compute the relative ranking of events (or clusters) by utilizing only the local parameters of a cluster without resorting to any global data structure or entity; since our objective is to discover events in real time any global computation (for instance, relative ranking of events by considering all the events in the current time window) for ranking is simply not scalable. Therefore, for efficient ranking of the clusters, we take into account *local* cluster properties, namely:

a) Correlation coefficient of edges present in a cluster.
b) Density of cluster (number of edges present in a cluster).
c) Support of the cluster, i.e., the number of independent user ids associated with the cluster keywords.

A natural way to think about these parameters is that a strongly correlated dense cluster with high support should be ranked higher. Hence, a set of messages due to a real event is more likely to be ranked higher than an accidental cluster formation as accidental clusters are likely to possess low correlation, low density, or low support.

Let $C = (V,E)$ is a cluster discovered by our algorithm. *V* is the set of nodes in C, $|V| = n$. *E* is set of edges in *C*. We compute the rank of the cluster as $\frac{1}{n}.\mathbf{W.C}$ where **W** is the weight matrix of size **1-by-n** where $w_i$ is the weight of a node *i*, i.e., the number of user ids associated with it. **C** is edge correlation coefficient matrix of size **n-by-n**. $C_{ii} = 1; \forall i, C_{ij} = 0; (i,j) \notin E$. We normalize the cluster rank with its size so that the rank of a cluster is not a monotonically increasing function of cluster size. Hence, a strongly connected cluster will result in higher rank as there would be many non-zero entries in **C**. Secondly, higher correlation coefficient values will result in higher cluster rank. Finally, higher support to cluster will result in higher value of weight matrix, **W,** resulting in higher rank.

## 7. EXPERIMENTAL EVALUATION
Our goal in Section 7.1 is to compare and contrast our SCP based technique, designed to extract, in real-time, emerging events from microblog messages, with ground truth regarding real-world events, as manifested in Google news headlines. The above study establishes that our technique is capable of identifying real-world events, as they occur, with high precision and recall. In the experiment reported in Section 7.2 we present the results of a detailed precision and recall study using Twitter traces. In Section 7.3, we compare the performance of our *SCP* based clustering algorithm with an offline method [2].

### 7.1 Evaluation against Ground Truth
Using an RSS feed reader we collected a total of 473 Google news headlines over a period of 18 hours on $29^{th}$ Feb 2012. 255 of these headlines related to USA specific real time events (for example, we did not consider news analysis related headlines among our events of interest). These headlines were found to be related to 60 unique real-world events. We concurrently ran a twitter downloader to obtain more than 1.3 million tweets generated within the USA (by providing longitude and latitude range). Tweet download rate was close to 21/sec.

We set δ=800 tweets/quantum, and *w=30 quanta,* representing a history of 20 minutes. Note that *δ* is defined in terms of number of messages in our experiments. First we identified all the bursty keywords in the twitter trace; a keyword is bursty if in at least one quantum in the trace, the keyword is used by ≥ 4 users. That is, *γ=4. λ*=0.1. A keyword in a given Google news headline, (after removing stop words), must be present in the bursty keyword list in order to be identified as pertaining to an event. For instance, corresponding to the headline "Body of missing Florida firefighter found", there was only one tweet present in the entire trace. *γ*=4 implies that the event represented by the lone tweet need not be considered as an emerging event. Of the 60 news events, there were 27 such events (with very few related tweets) including, "Egypt lifts travel ban on 7 US pro-democracy workers", "Rep. David Drier decides against seeking reelection", etc. Of the remaining 33 emerging real-world event related headlines our technique identified 31 events. Other two headlines were (("Obama, Congress leaders seek cooperation on jobs", "Obama praises Snowe"). Of the keywords occurring in these headlines, only "Obama" exceeded the burstiness threshold. Upon investigation, we found that considered w.r.t. each headline, "Obama" could not be characterized as being bursty. Hence our technique did not report these two events.

In the table below, we list a subset of the identified events. Events with real time implications such as weather warning (Tornado in MidWest) were up to 6 hours ahead of their Google News counterpart. Some events, like 'Apple', were concurrent with the source of Google news (USA Today).

**Table 1: *SCP* technique w.r.t. ground truth**

| Google News HeadLine | Event Discovered Using *SCP* |
|---|---|
| Davy Jones of Monkees dead | Davy Jones Monkees Dead RIP |
| Tornado pounds MidWest | Watch awesome Tornado outside |
| A dead body found by Miami police on Rick Ross's House | Dead body found Rick house |
| Nebraska senator Bob Kerrey reverses decision not to run | Bob Kerrey will run |
| Apple market value hits $500B | Apple worth more than Poland |

It is important to point out that we identified almost 6 times more events (e.g., "Forecast 29$^{th}$ Feb Snow Rain Today", "advisory high wind warning issued surf") which were not present in Google news headline but were important in the local context (see Table 3). These included local job openings such as "#jobs alert ca #job #retail store #accounting manager #tweetmyjobs".

## 7.2 Precision and Recall

### 7.2.1 Experimental Setup

For a more detailed study of our technique and to understand its sensitivity to various parameters, we used 2 different data sets, 1) Event Specific (*ES*), (comprising a total of 8 million tweets) containing tweets corresponding to specific topics such as the Japan earthquake, Apple, etc. 2) Time Window (*TW*), (comprising a total of 10 million tweets) contains tweets, generated during a particular time window, not specific to any event or location. Tweets appear in chronological sequence w.r.t. their time of generation. The tunable parameters are listed in Table 2.

**Table 2: Nominal values used in the experiments**

| Parameter Name | Nominal Value | Tunable Range |
|---|---|---|
| *Quantum size δ* | 160 tweets | 80-240 tweets |
| *HighStateThreshold γ* | 4 user ids/quantum | NA |
| *EC Threshold λ* | 0.20 | 0.1-0.25 |
| *Window Length w.δ* | 30 quanta | 20-40 quanta |

Window length is 30 quanta, comprising a total of 4800 most recent tweets. The events in our case comprise global news such as "Plane crash in Iran kills 150 passengers" to more specific or local events such as "Now milk products in Fukushima are contaminated". Many a times events may not be breaking news for world media but important in the local context.

Identification of an event depends on the nodes and edges that constitute the graph. Therefore, in our experiments, we have varied two parameters to test our algorithm's performance; 1) Quantum size ($δ$). $δ$ is related to the burstiness of keywords. The larger the $δ$, the less bursty an event needs to be and vice versa; 2) *Edge Correlation threshold* ($λ$). We report the results obtained by varying the quantum size instead of *high state threshold* ($γ$) as, if we vary $γ$ the set of events itself changes. It is important to point out that varying $γ$ shows similar trends.

### 7.2.2 Measuring Recall and Precision

If a trace contains messages pertaining to an event but we do not discover the event, *loss of recall* occurs. We may miss the event due to (1) non-formation of the corresponding cluster (i.e., only 1 or 2 words from the event showed burstiness) (2) the cluster formed does not satisfy the *SCP*. Therefore, we compute recall as follows: First we collect all the keywords, after removing stop words, which are either bursty (based on the *high state* threshold $γ$) or are already present in the current sliding window. An event in the current window can comprise of only these keywords. Keywords which are bursty but not present in any of the chronologically correlated event clusters discovered in offline manner indicate potentially missed events. Once we collect all such *noun* words (we use Stanford POS Tagger [16]), we manually check in the trace if they indeed belong to any real event or not. To make this check manageable, given the size of the data, we randomly pick a fraction of missing noun keywords. The probability of them belonging to a real event is extrapolated to estimate the number of missing events. The maximum number of events (by adding both events identified and events missed by our algorithm) discovered in a run is considered to be the sum total of all the events present in the trace. We use this number to compute recall across different runs. Once the maximum number of real events is estimated, the same number is used to compute recall across all the runs. Therefore, our objective of studying the impact of parameter tuning on recall is not affected because of 'estimation' inaccuracy, if any.

Precision is defined as 'How many of the events identified by us are real and important events?' A spurious event reported by our system leads to *loss of precision*. However, for an event, classification of it as real or spurious can be subjective. Therefore, to identify spurious events, we employ the following approach: 1) We ignore an event if its rank is below a threshold which is a function of the minimum rank that a cluster of size $N$ can have (for given correlation and burstiness thresholds); 2) we ignore the clusters with all non-noun words. Our premise is that there must be at least one noun keyword in real world events.

However, there may still be spurious event clusters (such as advertisements or rumors). As noticed in our evaluation, real world events typically have a build-up and wind-down phase. Therefore, the clusters belonging to such events are evolving and/or their rank scores keep on changing in a non-monotonic manner. On the other hand, spurious events have a sudden burst and thereafter they die. Hence, events which do not evolve and have monotonically decreasing rank scores are considered spurious events in our analysis. We cannot suppress these events from being reported as we cannot determine their future behavior. However, for an event, we can analyze its behavior in a post-hoc manner. Precision for events is computed as the percentage of real events among all the events that are reported.

### 7.2.3 Observed Precision and Recall

The event density (events/unit length of trace) in ES set is found to be approximately 3 times that in TW set. Recall and Precision results are shown in Figures 7 to 10. In general with increasing $δ$ and decreasing $λ$, recall increases as more nodes and edges move into AKG due to the less stringent requirement on the burstiness. Similarly, precision tends to improve with increasing $δ$ and decreasing λ (though not as much as recall) due to the following reason: in our experiments we see that spurious events tend to appear in bursts. Hence, there is practically no effect of parameter tuning on these events due to their strong temporal correlation and they are almost always discovered in each run. However, with more relaxed parameters, majority of the *extra* events that get discovered are real events. Hence, with more events getting identified and the number of spurious events remaining approximately stable, precision increases. In experiment (run on ES trace) with δ=800, λ=0.25 and γ=4: (i) Recall improves to 0.95; (ii) Precision also improves marginally due to the presence of almost the same number of spurious events. As stated earlier, we varied δ instead of γ to see the effect of burstiness. Finally, changing *w*, the number of quanta in a window, did not result in a discernable effect on precision/recall.

### 7.2.4 Analysis of Quality of Discovered Events

From our previous experiment, it may appear that one may set the δ as large and λ as small as possible to achieve maximum recall and precision. However, another important dimension in our analysis is the quality of the discovered events. With low λ and high δ, more and more keywords start merging with event clusters, reducing the quality of event clusters. Similarly, many meaningless (or less interesting) events may get discovered. We use the following two measures to determine the event quality: 1) *Average cluster size:* We compute the size of average cluster for all the discovered events. The average cluster size across all the runs ranges from 6.16 to 6.88 keywords/event except when λ is reduced to 0.1 when the average cluster size becomes 9.23 and 9.88 for ES and TW data sets respectively indicating an almost 50% increase. As one can see, consuming small and focused event clusters is preferred compared to large clusters. 2) *Average Cluster Rank:* As explained in Section 6, a high rank score signifies a strong cluster and therefore a better event quality. We notice, with increasing δ and reducing λ, average rank score reduces by up to 20% and 30% in TW and ES traces respectively from its peak value. As we see, the average cluster size does not change much across different runs, the reduced rank score implies that most of the additional events that are discovered with more relaxed parameters have fairly low rank score. Further, clusters around real events were almost always ranked higher compared to clusters formed accidentally.

## 7.3 Bi-connected clusters vs. SCP clusters

We implemented the algorithm to discover the bi-connected clusters (*BCs*) on exactly the same graph on which *SCP* clusters are computed. Similar algorithm is also proposed in [2] to identify events in blogs. After each quantum, the *BCs* are computed on the entire graph in an offline manner. All the edges (including edges connecting two *BCs*), which are not part of any bi-connected cluster, are reported as clusters of size 2. All parameters are set to their nominal values (Table 2). We have used the same twitter trace which we used for the ground truth experiment. At the end of each quantum, clusters identified by both the techniques are compared. Since *SCP* is not a necessary condition for bi-connected clusters, additional clusters are discovered in the offline method. Therefore, we compute: (1) additional clusters ($A_c$) and (2) additional events ($A_E$) discovered in offline method. We get 276% $A_c$ and -11.1% $A_E$. If we exclude BCs of size 2 from offline clusters (since *SCP* based clusters do not include them), $A_c$ and $A_E$ come down by -5.1% and -17.1%. The additional clusters in the offline method arise from edges being identified as clusters (of size 2). A substantial number of these edges are found to be not related to *real* events. We further identify that 1) 74.5% of offline event clusters exactly overlap with *short-cycle* based clusters (after excluding edges), 2) no instance of an event cluster is found in the offline method which did not have *short-cycle*. Both these facts prove (1) the correctness of our method (2) our conjecture that *real* events have *short-cycle* within the event cluster. Average size of exactly overlapping clusters was 4.53 (against 5.07 for all the clusters in the *SCP* method) indicating that mostly small clusters overlap exactly. For the offline event clusters, not overlapping *exactly* with *short-cycle* based clusters, we see an increase in average cluster size from 6.83 to 12.72. The average rank of all the *BC* clusters goes down from 186.4 to 150.9 w.r.t. SCP clusters. Therefore, the quality of offline clusters suffers. Further, our technique computes clusters 46% faster compared to offline method due to the fact that it involved *only* local computations. Note that the performance of our method can further be improved in a parallel processing environment since multiple simultaneous computations are allowed on the graph in *short-cycle* based clusters.

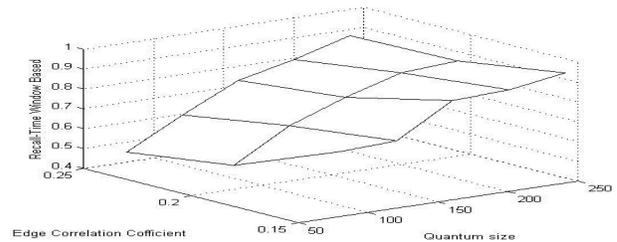

**Figure 7: Recall for Time Window Based Trace**

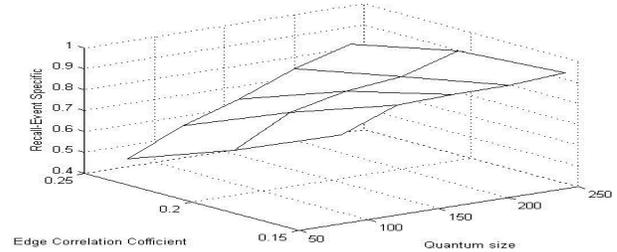

**Figure 8: Recall for Event Specific Trace**

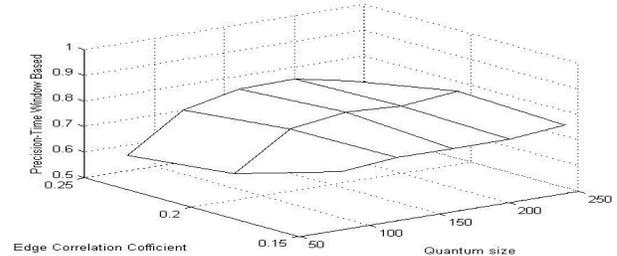

**Figure 9: Precision for Time Window Based Trace**

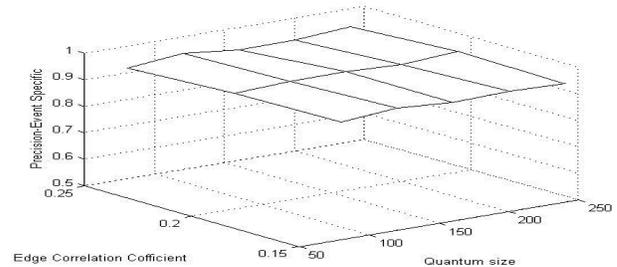

**Figure 10: Precision for Event Specific Trace**

As is evident from the above discussion, the offline clusters lead to lower precision. However, even recall is lower as in some instances two *real* events get merged into one offline cluster, leading to loss of recall. In summary, we show that *real* time events almost invariably have *short-cycle* within the cluster (except a small number of events which do not form cycles).

**Table 3: Performance of different clustering schemes**

|  | *SCP Clusters* | *Bi-connected Clusters* | *Bi-connected clusters +Edges* |
|---|---|---|---|
| Events Discovered | 216 | 179 | 192 |
| Precision | 0.911 | 0.795 | 0.216 |
| Recall | 0.935 | 0.775 | 0.831 |
| Avg. Rank | 186.4 | 150.9 | 92.1 |
| Avg. Cluster Size | 5.07 | 6.31 | 3.14 |

## 7.4 Impact of using AKG

Recall from Section 3 that at the end of each quantum, we do a total of $O(k^2NC)$ computations. On average, the number of edges in AKG was less than 2% of CKG (at a given point of time). In our experiments, less than 5% nodes in CKG show burstiness. These reductions demonstrate the efficacy of our technique to reduce the size of graphs used for cluster discovery.

Further, the average number of edges attached to a node was less than 6 and the average size of clusters was less than 7 nodes. Hence we can clearly see that the amount of computation that needs to be done at the end of each quantum is significantly reduced due to the use of SCP over AKG. In the table below, we show the message processing rate. We see that on our machine, one with modest configuration, the rate of processing a general twitter trace is beyond 5000 messages/second. On *ES* trace, with much higher event intensity, the rate of processing comes down. With increasing δ, the number of low quality clusters increases and only some of them are identified as real events. The system ends up processing many clusters which are discarded later.

**Table 4: Message processing rate for given quantum sizes**

| Trace Type | Msg Processed/Second | | |
|---|---|---|---|
|  | δ =120 | δ =160 | δ =200 |
| Time Window Based Trace | 5185 | 4420 | 4160 |
| Event Specific Trace | 1410 | 1400 | 1160 |

In summary, our experiments demonstrate the following:
- We see that our algorithm is able to discover interesting events with high precision and recall in a timely manner. Besides 'important' events it also discovers events which may not be "captured" by headlines reported in news sites.
- Our algorithm, by exploiting the SCP, works in real time and outperforms the offline algorithm reported in [2]. Analysis of events identified in offline method also establishes that SCP is almost invariably present in all the event clusters.
- Our algorithm is quite resilient to parameter settings as the event set discovered by us is fairly stable across different runs underlining the robustness of our algorithm. Further, we find that the average cluster size is quite stable across runs.
- On a modest machine, our algorithm is able to process almost twice the current rate at which messages are added to the Twitter stream underlying our algorithms' scalability.

## 8. CONCLUSION

In this paper we have addressed the problem of discovering events in a microblog stream. We mapped the problem of finding events to that of finding clusters in a graph. Due to the dynamic nature of the twitter stream, the size of the graph can become extremely large. We hence proposed the use of a technique which allowed us to efficiently find a stable graph which was order of magnitude smaller than the original graph and yet captures all the information about the emerging events. We argued that conventional cluster discovery techniques used for finding events in a microblog stream do not work in our setting. We hence introduced *aMQCs*, which are bi-connected clusters, satisfying a new *short-cycle property* which allowed us to find and maintain the clusters locally without affecting the quality of the discovered clusters. To handle the dynamics we also proposed algorithms for handling addition/deletion of a node/edge and proved the correctness of the same. We showed the efficacy of our techniques using real world data – we were able to find clusters efficiently, in real-time, i.e., keeping pace with the arrival of messages.

As part of our future work we plan to build a system that includes the ability to discover and thus discard spurious and malicious events (e.g., rumors). Since many web applications generate data which can be modeled as massive and dynamic graphs, we will also extend and apply our technique to other domains with similar characteristics. Finally, we will explore pre- as well as post-processing techniques to complement the core approach described in this paper.